\documentclass[a4paper,11pt]{article}
\usepackage{jheppub2}
\usepackage[utf8x]{inputenc}
\usepackage{graphicx}
\usepackage{float}
\usepackage[T1]{fontenc}
\usepackage{epsfig}
\usepackage{color}
\usepackage{tikz}
\usepackage{amsmath,mathtools}
\usepackage{subfloat}
\usepackage{amsfonts}
\usepackage{braket}
\usepackage{cleveref}
\usepackage{epstopdf}
\usepackage{caption}
\usepackage{subcaption}
\usepackage{enumitem}
\usepackage{etoolbox}
\usepackage{orcidlink}
\usepackage{comment}
\graphicspath{ {./Figures/} }
\usepackage[titletoc,toc,title]{appendix}
\usepackage{hyperref}
\hypersetup{
	colorlinks=true,
	linkcolor=blue,
	filecolor=red,      
	urlcolor=blue,
	citecolor=red
} 
\usepackage{tikz}
    \usepackage{pgfplots}
    \pgfplotsset{compat=1.17} 
\usepgfplotslibrary{fillbetween}
\usetikzlibrary{positioning,decorations.pathmorphing}

\title{Universal Supercritical Behavior in Global Monopole-Charged AdS Black Holes}

\author[a]{Ankit Anand\orcidlink{0000-0002-8832-3212}}
\author[b,*]{\;and\;Shoucheng Wang\orcidlink{0000-0002-5909-6233}}

\affiliation[a]{
	Department of Physics, Indian Institute of Technology Kanpur, 208016, India}
    \affiliation[b]{
	School of Science, Hunan Institute of Technology, Hengyang 421002, China.}

\emailAdd{anand@iitk.ac.in}
\emailAdd{scwang@hnit.edu.cn}

\let\oldthefootnote\thefootnote
\renewcommand{\thefootnote}{}
\footnotetext{*Corresponding author.}
\let\thefootnote\oldthefootnote

\date{}

\abstract{ We analytically investigate the Widom line and universal supercritical crossover for charged AdS black holes threaded by a global monopole. We compute thermodynamic variables in both the extended and canonical ensembles. We derive the scaled variance $\Omega$ using the Gibbs free energy and locate the Widom line as the extrema of this. Using mean-field expansion of the equation of state near criticality, we obtain closed-form expressions for the Widom line and the two branching crossover lines $L^\pm$. We show that the monopole parameter shifts the critical parameters but does not change the mean-field universal scaling: the leading linear term and the nonanalytic correction remain universal in both ensembles. We also numerically verify this using the supercritical crossover lines $L^\pm$ and show the universal scaling laws and the complete supercritical phase diagrams.
}

\begin{document}
	
\maketitle


\newpage

\section{Introduction}

First-order liquid-gas phase transitions terminate at a critical point, beyond which thermodynamic distinctions between the liquid and gas phases become continuous in ordinary fluids. In this supercritical region, no sharp phase boundary exists, yet the system exhibits pronounced thermodynamic crossovers. These crossovers are characterized by the Widom line~\cite{widom_line_ref, stanley1971phase, Ruppeiner:2011gm}, defined as the locus of maxima in response functions such as the specific heat, compressibility, or thermal expansion coefficient. The Widom line provides a natural extension of the liquid-gas coexistence curve into the supercritical domain, delineating regions that behave like liquids and those that behave like gases in a continuous manner.

A similar structure appears in black hole thermodynamics because numerous AdS black hole solutions display van der Waals-type behavior~\cite{Kubiznak:2012wp, Gunasekaran:2012dq, Chamblin:1999tk, Wei:2023mxw, Spallucci:2013osa}, where first-order phase transitions terminate at the critical point. Beyond this point, black holes enter into a supercritical regime in which thermodynamic observables exhibit smooth but pronounced extrema. These features enable the definition of a black hole Widom line~\cite{Xu:2005widom, Simeoni:2010widom}, which delineates distinct supercritical regimes~\cite{Ouyang:2024ckt, Xu:2025jrk} and closely parallels the behavior observed in ordinary fluids. Near criticality, the scaling of response functions follows mean-field behavior, reflecting universality classes shared across disparate physical systems~\cite{stanley1971phase, ma1976modern}. The study of crossover phenomena, therefore, provides a powerful pathway for probing the microscopic and universal properties of black hole thermodynamics, offering insight into how critical fluctuations persist even in the absence of a true phase transition. In the supercritical regime, black holes no longer exhibit a sharp first-order transition; yet, thermodynamic observables retain strong remnants of criticality~\cite{Li:2025lrq, Zhao:2025ecg}.

Black holes may coexist with relics of early-Universe symmetry breaking, and such defects can imprint geometric modifications on the spacetime. Among these, global monopoles provide a particularly clean and analytically tractable example, introducing a solid-angle deficit without altering the basic topology of the horizon. Topological defects provide a complementary arena for exploring modified gravitational and thermodynamic behavior. Global monopoles arise naturally when a global $O(3)$ symmetry is spontaneously broken to $U(1)$ during early-Universe phase transitions~\cite{Kibble:1976sj, Vilenkin:1984ib}. Their long-range Goldstone fields generate an energy density scaling as $\rho(r)\sim r^{-2}$, leading to a finite solid-angle deficit rather than a localized gravitational potential~\cite{Barriola:1989hx, Shi:1991yto}. When embedded in or absorbed by black holes, global monopoles introduce an additional topological charge that modifies the spacetime geometry and the associated thermodynamic variables~\cite{Yu:1994fy, Ahmed:2016ucs,Tan:2017egu}. In AdS spacetimes, the monopole parameter rescales key thermodynamic quantities, thereby enriching black hole chemistry~\cite{Kubiznak:2014zwa} and influencing the microscopic degrees of freedom associated with the horizon.

Although charged AdS black holes with a global monopole~\cite{Banerjee:1996uk, Jing:1993np, Yu:1994fy, Dadhich:1997mh, Li:2002ku} have been shown to have similar Van der Waals-like criticality as other AdS black holes, prior work has focused on equilibrium coexistence and critical points~\cite{Deng:2018wrd, Soroushfar:2020wch, Luo:2022gss, Cui:2024xyl}. The supercritical crossover structure, in particular, the analytic determination of the Widom line and the bifurcating crossover branches $L^{\pm}$ in the presence of a monopole, remains unexplored. This paper fills that gap by deriving closed-form expressions for the scaled variance $\Omega$, the Widom line in both ensembles, and the universal mean-field corrections, thereby quantifying how a topological solid-angle deficit reshapes near-critical thermodynamics without altering mean-field universality. Overall, the study of the Widom line and supercritical crossover in global monopole spacetimes reveals how geometric and topological modifications influence universal thermodynamic behavior, extending the fluid–gravity analogy into new territory and deepening the connection between phase transitions in gravitational and condensed matter systems.

In this work, we perform an analytical investigation of the Widom line and the associated universal crossover behavior in charged AdS black holes with a global monopole. Using the mean-field approach, we determine how the monopole parameter reshapes both the critical behavior and the supercritical thermodynamic structure. The rest of the paper is organized as follows. In Section~\ref{Sec:Review of Known Results}, we review the known results starting with the Black Hole Solution with their thermodynamics in the Extended Ensemble~\ref{Sec:ExtendedEnsemble}, the Canonical Ensemble~\ref{Sec:CanonicalEnsemble} and the Mean-field description and the role of susceptibility in black hole criticality~\ref{Subsec:Mean-field description and the role of susceptibility in black hole criticality}. In Section~\ref{Sec:Widom Line Computation via Variance}, we present the Widom Line Computation via Variance and derive the Widom temperature and confirm the universal mean-field behavior. In Section~\ref{Sec:Supercritical Crossover}, we derive the supercritical crossover lines analytically and verify them numerically. The corresponding phase diagrams emphasize a clear analogy between the supercritical behavior of black holes and water systems. Finally, Section~\ref{Sec:Conclusion} summarizes our results.


\section{Review of Known Results}\label{Sec:Review of Known Results}

Global monopoles arise naturally in scalar field theories exhibiting spontaneous symmetry breaking from a global $O(3)$ symmetry to $U(1)$~\cite{Barriola:1989hx}. Their long-range Goldstone fields modify the asymptotic structure of spacetime, producing a solid-angle deficit rather than a localized potential. When embedded in an AdS background, this geometric deformation alters both the horizon geometry and the associated thermodynamic phase space. The resulting system provides a tractable setting for studying how symmetry-breaking topological defects influence black hole criticality. Moreover, in the context of the AdS/CFT correspondence, such deformations may encode nontrivial vacuum structures or the presence of topological defects in the dual field theory. In this section, we review the construction of a charged AdS black hole carrying a global monopole and summarize its thermodynamic behavior in both the extended and canonical ensembles. These results will form the foundation for our later analysis of the supercritical regime and the emergence of the Widom line.

\subsection{Black Hole Solution}

The simplest model generating a global monopole consists of a triplet scalar field $\varphi^a$ $(a = 1,2,3)$ with a global $O(3)$ symmetry broken to $U(1)$ as
\begin{equation}
\mathcal{L} = \mathcal{R} - 2\Lambda 
+ \frac{1}{2}\,\partial_\mu \varphi^a \partial^\mu \varphi^{*a}
- \frac{\gamma}{4}\left(\varphi^a \varphi^{*a} - \zeta_0^2\right)^2 ,
\end{equation}
where $\gamma$ is the self-coupling and $\zeta_0$ fixes the symmetry-breaking scale. The scalar configuration takes the hedgehog form
\begin{equation}
\varphi^a = \zeta_0\, b(\bar{r})\, \frac{\bar{x}^a}{\bar{r}} \ , \qquad  \bar{x}^a \bar{x}^a = \bar{r}^2,
\end{equation}
with $b(\bar{r}) \to 1$ as $\bar{r} \to \infty$. Outside the monopole core, the energy-momentum tensor becomes effectively fixed and induces a deficit solid angle~\cite{Barriola:1989hx}, which is responsible for the characteristic modification of the geometry.

To construct a charged AdS black hole with a global monopole, one adds the monopole sector to the Einstein-Maxwell action. The resulting static, spherically symmetric solution~\cite{Rhie:1990kc, Deng:2018wrd, Soroushfar:2020wch} is
\begin{equation}
ds^2 = -\left(1 - 8\pi\zeta_0^2 - \frac{2\bar{m}}{\bar{r}} + \frac{\bar{q}^2}{\bar{r}^2} + \frac{\bar{r}^2}{L^2} \right)d\bar{t}^2+ \frac{d\bar{r}^2}{1 - 8\pi\zeta_0^2 - \frac{2\bar{m}}{\bar{r}}+ \frac{\bar{q}^2}{\bar{r}^2} + \frac{\bar{r}^2}{L^2}}+ \bar{r}^2 d\Omega_2^2 \ .
\end{equation}
where $L$, $\tilde m$, and $\tilde q$ denote the AdS radius, the mass parameter, and the electric charge parameter of the black hole, respectively. The AdS radius is related to the cosmological constant by $\ell^2=-3/\Lambda$ (with $\Lambda<0$). The parameter $8\pi\zeta_0^2$ encodes the solid-angle deficit. Because of this metric, thermodynamic quantities of the black hole display a nontrivial dependence on the global monopole parameter, and they satisfy both the first law of thermodynamics and the Smarr relation. We introduce the rescalings and coordinate redefinition~\cite{Barriola:1989hx} as 
\begin{eqnarray}
    \bar{t} &=& \frac{t}{\sqrt{1-\zeta^2}} \ , 
\qquad \bar{r} = (1-\zeta^2)^{1/2} r \ ,\nonumber \\[2mm]
M &=& \frac{\bar{m}}{\sqrt{1-\zeta^2}}, 
\qquad Q = \frac{\bar{q}}{\sqrt{1-\zeta^2}} \ , \\[2mm]
&&\qquad\qquad \zeta^2 \equiv 8\pi\zeta_0^2 \ . \nonumber
\end{eqnarray}
With this, the metric takes the more convenient form as
\begin{equation}
ds^2 = -f(r)\,dt^2 + \frac{dr^2}{f(r)} + r^2(1-\zeta^2)\, d\Omega_2^2,
\end{equation}
with 
\[
f(r)=1 - \frac{2M}{(1-\zeta^2)r}
+ \frac{Q^2}{(1-\zeta^2)^2 r^2}
+ \frac{r^2}{L^2} .
\]

This rescaled geometry makes explicit that the monopole induces a multiplicative deficit in the angular sector while leaving the radial and temporal structure unchanged up to parameter redefinitions. This feature plays a central role in the thermodynamic modifications discussed below.

\subsubsection{Thermodynamics in the Extended Ensemble}\label{Sec:ExtendedEnsemble}

In the extended phase space, the cosmological constant is promoted to a thermodynamic variable interpreted as pressure~\cite{Kastor:2009wy,Dolan:2011cs,Cvetic:2010jb,Dolan:2011xt,Kubiznak:2014zwa},
\begin{equation}
P = -\frac{\Lambda}{8\pi} = \frac{3}{8\pi L^2}.
\end{equation}
The black hole mass $M$ corresponds to enthalpy, and the first law and Smarr relation take the form
\begin{equation}
dM = T\,dS + V\,dP + \Phi\,dQ,
\qquad
M = 2TS + \Phi Q - 2PV.
\end{equation}

The thermodynamic quantities expressed in terms of the horizon radius \( r_+ \) are
\begin{eqnarray}\label{Temp_relation}
T &=& \frac{1}{4\pi r_+}
\left(1 + \frac{3 r_+^2}{L^2}
- \frac{Q^2}{(1-\zeta^2)^2 r_+^2}\right), \\
S &=& \pi(1-\zeta^2) r_+^2, \qquad
V = \frac{4\pi}{3}(1-\zeta^2) r_+^3, \qquad
\Phi = \frac{Q}{(1-\zeta^2) r_+}.
\end{eqnarray}

Substituting these into the equation of state yields
\begin{equation}\label{Main_EOS}
P = \frac{T}{2 r_+} - \frac{1}{8\pi r_+^2}
+ \frac{Q^2}{8\pi (1-\zeta^2)^2 r_+^4}.
\end{equation}
The monopole parameter modifies the small/large black hole phase transition by shifting the critical point \((P_c, r_{+c}, T_c)\). Solving the inflection-point conditions
\[
\partial_{r_+}P|_{T}=0, \qquad 
\partial_{r_+}^2 P|_{T}=0,
\]
one obtains
\begin{equation}\label{P_cv_CT_c}
v_c=\frac{2\sqrt{6}\,Q}{1-\zeta^2}, \qquad
T_c = \frac{1-\zeta^2}{3\sqrt{6}\,\pi Q}, \qquad
P_c = \frac{(1-\zeta^2)^2}{96\pi Q^2}.
\end{equation}
Although each critical quantity is shifted by the monopole, the universal ratio
\[
\frac{P_c v_c}{T_c}=\frac{3}{8}
\]
remains unchanged, confirming that the system retains mean-field Van der Waals universality. This ensemble therefore provides a natural framework for analyzing black hole $P$--$v$ criticality, scaling behavior, and the influence of the monopole on the near-critical phase structure.

\subsubsection{Thermodynamics in the Canonical Ensemble}\label{Sec:CanonicalEnsemble}

In the canonical ensemble, the AdS radius—and hence the pressure—is held fixed, while the charge $Q$ is taken as the fluctuating variable. The appropriate potential is the Helmholtz free energy,
\begin{equation}
F(T,Q)=M-TS,
\end{equation}
with the first law
\begin{equation}
dF = -S\,dT + \Phi\,dQ.
\end{equation}

Expressed in terms of the entropy and charge, the Hawking temperature becomes
\begin{equation}\label{Temp_in_S_Q}
T(S,Q)= \frac{\pi L^2\left[(1-\zeta^2)S - \pi Q^2\right] + 3S^2}{4\pi^{3/2}\sqrt{1-\zeta^2}\,L^2\, S^{3/2}}.
\end{equation}
Local thermodynamic stability is governed by the heat capacity,
\[
C_Q = T\left(\frac{\partial S}{\partial T}\right)_Q,
\]
while global stability is reflected in the Helmholtz free energy. The critical point is determined by
\[
\left(\frac{\partial T}{\partial S}\right)_Q = 0,
\qquad
\left(\frac{\partial^2 T}{\partial S^2}\right)_Q = 0,
\]
yielding
\begin{equation}\label{S_cQ_cT_c}
S_c = \frac{\pi (1-\zeta^2) L^2}{6},
\qquad
Q_c = \frac{(1-\zeta^2)L}{6},
\qquad
T_c = \sqrt{\frac{2}{3}}\frac{1}{\pi L}.
\end{equation}

The canonical ensemble thus complements the extended ensemble by highlighting how the global monopole reshapes the charge sector of the thermodynamic landscape while preserving the underlying mean-field critical behavior. Together, these two thermodynamic descriptions provide a coherent picture of the monopole-induced modifications to the phase structure, preparing the ground for our subsequent analysis of supercritical crossovers and the Widom line.


\subsection{Mean-field description and the role of susceptibility in black hole criticality}\label{Subsec:Mean-field description and the role of susceptibility in black hole criticality}

The mean-field approach provides a universal framework to describe critical phenomena across a wide range of systems, from ordinary fluids and magnets to black holes~\cite{Stanley1971,Goldenfeld1992}. In this framework, the essential features of a phase transition are captured by introducing an \emph{order parameter} $\phi$, which quantifies deviations from the symmetric (disordered) phase. Near the critical point, the thermodynamic potential can be expanded as a power series in $\phi$, giving the Landau free-energy functional
\begin{equation}
\mathcal{F}(\phi,T) = \mathcal{F}_0(T) + \frac{1}{2}a(T)\phi^2 + \frac{1}{3}b\phi^3 + \frac{1}{4}c\phi^4 - h\phi + \cdots,
\end{equation}
where $a(T)$ changes sign at the critical temperature $T_c$, $b$ and $c$ are temperature-independent coefficients characterizing nonlinear interactions, and $h$ is the external field conjugate to $\phi$.

The equilibrium configuration minimizes $\mathcal{F}$, leading to the mean-field equation of state,
\begin{equation}
\frac{\partial \mathcal{F}}{\partial \phi} = 0
\quad \Rightarrow \quad
a(T)\phi + b\phi^2 + c\phi^3 - h = 0 .
\end{equation}
This relation encodes the nonlinear response of the system to its conjugate field and provides the foundation for defining the \emph{susceptibility},
\begin{equation}
\chi = \left(\frac{\partial \phi}{\partial h}\right)_T,
\end{equation}
which diverges near the critical point as
\begin{equation}
\chi \sim |\tau|^{-\gamma}, \qquad \tau = \frac{T-T_c}{T_c}, \quad \gamma = 1 \ (\text{mean-field}).
\end{equation}
The divergence of $\chi$ signals the emergence of long-range correlations and the critical amplification of fluctuations. In the supercritical regime, this enhanced response persists as a smooth crossover, giving rise to the \emph{Widom line}, which represents a continuous extension of critical ehavior beyond the phase coexistence region.

A salient feature of mean-field theory is \emph{universality}: the critical exponents depend only on the symmetry of the order parameter and the dimensionality of the system, independent of microscopic details~\cite{Fisher1974}. The classical mean-field exponents are
\begin{equation}
\alpha = 0, \quad \beta = \frac{1}{2}, \quad \gamma = 1, \quad \delta = 3,
\end{equation}
where $\beta$ governs the onset of $\phi$ below $T_c$, $\gamma$ characterizes the divergence of $\chi$, and $\delta$ determines the nonlinear response along the critical isotherm. These exponents provide a unified description of scaling near criticality in both gravitational and non-gravitational systems.

In black hole thermodynamics, the Widom line offers a natural generalization of these ideas to the supercritical regime. While the susceptibility $\chi$ quantifies the response at the critical point, the \emph{scaled variance} $\Omega$ serves as a generalized susceptibility in the one-phase region~\cite{Zhao:2025ecg}. Derived from second derivatives of the Gibbs free energy, $\Omega$ measures the strength of thermodynamic fluctuations of the order parameter—such as horizon radius, entropy, or electric potential—depending on the ensemble. Its maxima trace the Widom line, marking the locus of strongest crossover response between small- and large-black-hole–like states.  

The identification of the order parameter and its conjugate field depends on the chosen ensemble. In the extended phase space, the thermodynamic volume $V$ plays the role of $\phi$, with the isothermal compressibility
\begin{equation}
\kappa_T = -\frac{1}{V}\left(\frac{\partial V}{\partial P}\right)_T \ ,
\end{equation}
acting as the susceptibility $\chi$, diverging at the critical point~\cite{Dolan:2011cs}. In the canonical ensemble, the order parameter may be taken as deviations in the horizon radius $r_+$ or electric potential $\Phi$, with the specific heat $C_Q$ serving as an effective susceptibility, $\chi \sim (\partial \phi/\partial T)_Q$~\cite{Kubiznak:2012wp}. Analytically, the Landau potential predicts a bifurcation into two crossover lines, $L^\pm$, emanating from the critical point. In black holes, these correspond to the loci of enhanced specific heat or compressibility, all governed by the same universal scaling relations dictated by mean-field exponents. This demonstrates that the universality of critical dynamics is dictated by the structure of the thermodynamic potential rather than the microscopic details of the gravitational system, offering a predictive framework for locating and characterizing the Widom line.  

Consequently, the mean-field approach, combined with susceptibility and scaled variance diagnostics, provides a coherent and quantitative description of black hole criticality and the supercritical crossover, laying the groundwork for our subsequent analytic exploration of the Widom line in global monopole spacetimes.


\section{Widom Line Computation via Variance}\label{Sec:Widom Line Computation via Variance}

The Widom line is the natural thermodynamic continuation of the coexistence curve into the supercritical region, delineating the smooth crossover between distinct black hole phases even in the absence of a true first-order transition. In this regime, the discontinuous jump characteristic of the subcritical phase transition is replaced by a rapid but continuous change in thermodynamic observables, reflecting the persistence of critical correlations for $T > T_c$.

The Widom line is conventionally defined as the locus of maxima of response functions such as specific heat or isothermal compressibility in the supercritical region~\cite{Ouyang:2024ckt,Zhao:2025ecg}. In this work, following the approach of Ref.~\cite{Zhao:2025ecg}, we construct the scaled variance $\Omega$ from higher-order derivatives of the Gibbs free energy and identify its maxima as the Widom line. This method originates from the statistical description of enhanced thermodynamic fluctuations in the supercritical regime. Specifically,
the condition
\[
\left(\frac{\partial \Omega}{\partial T}\right)_P = 0
\]
identifies the locus of maximum thermodynamic response at fixed supercritical pressure, thereby defining the Widom line $T_W(P)$. This curve separates liquid-like and gas-like black hole microstructures, providing a continuous extension of critical dynamics into the supercritical domain. Its identification is crucial for understanding universality, scaling structure, and the influence of the global monopole on near-critical behavior.

It is worth noting that Ref.~\cite{Xu:2025jrk} proposed a definition of the Widom line based on the projection of complex Lee-Yang zeros onto the real phase plane. Although mathematically more abstract, this definition is physically equivalent to the response-function-extremum approach used here: both characterize the locus of enhanced thermodynamic response or incipient phase separation in the supercritical region. In fact, near the critical point, the projection of Lee-Yang zeros coincides with the maxima of response functions, providing theoretical support for the definition adopted in this work.

To study the Widom line in black holes, we begin with the Gibbs free energy $G = M - TS$, written in terms of the horizon radius $r_+$. The $n$th-order temperature derivative of Gibbs free energy at fixed pressure is defined as
\begin{equation}\label{K_n_def}
k_n \equiv \left(\frac{\partial^n G}{\partial T^n}\right)_P .
\end{equation}
Because $G$ and $T$ both depend on $r_+$, the derivatives may be evaluated using the chain rule, yielding
\begin{eqnarray}
k_1 &=& \frac{G'}{T'}  \qquad;\qquad  k_2 = \frac{G'' T' - G' T''}{(T')^3} \ .
\end{eqnarray}
The normalized variance then becomes
\begin{eqnarray}
    \Omega =\frac{k_2}{k_1} = \frac{G''}{G'T'}-\frac{ T''}{(T')^2}  \ .
\end{eqnarray}
which quantifies the strength of thermodynamic fluctuations. Maxima of $\Omega$ trace out the Widom line~\cite{Zhao:2025ecg,PhysRevE.85.031203}, marking the supercritical boundary across which response functions show pronounced but nonsingular variation. 

For the global monopole case, the normalized variance takes the form 
    \begin{eqnarray}\label{eq:Variance}
\Omega = \frac{8\pi(1-\zeta^2)^2 r_+^3}{(1-\zeta^2)^2 r_+^2 (8\pi P r_+^2 - 1) + 3 Q^2} \ .
\end{eqnarray}
The global monopole parameter $\zeta$ modifies $\Omega$ through its appearance in all thermodynamic quantities, thereby affecting both the location and the shape of the Widom curve. As expected, it $\Omega$ diverges at the critical point, confirming that the Widom line originates at the critical endpoint. In the next subsection, we compute the Widom line in both ensembles.

\subsection{Widom line in the Extended ensemble}
To determine the Widom line explicitly, we impose the extremum condition $\partial_{r_+}\Omega = 0$, which leads to
\begin{equation}
8\pi (1-\zeta^2)^2 P\, r_+^4 + (1-\zeta^2)^2 r_+^2 - 9 Q^2 = 0 .
\end{equation}
The physical (positive) solution for the horizon radius is
\begin{equation}\label{r_W2inP}
r_W^2 =
\frac{\sqrt{(1-\zeta^2)^4 + 288\pi (1-\zeta^2)^2 P Q^2} - (1-\zeta^2)^2}{
16\pi (1-\zeta^2)^2 P} .
\end{equation}
Introducing the reduced pressure deviation $\Delta P = \frac{P - P_c}{P_c}$, the expression becomes
\begin{equation}\label{r_W2inDeltaP}
r_W^2(\Delta P)
= \frac{6Q^2 \left(\sqrt{3\Delta P + 4} - 1\right)}
{(1-\zeta^2)^2(\Delta P + 1)} .
\end{equation}
Substituting this into the Hawking temperature~\eqref{Temp_relation} gives the Widom temperature,
\begin{equation}
T_W(\Delta P)
= \frac{Q^2}{
3\sqrt{6}\pi(1-\zeta^2)^2
\left(
\frac{(\sqrt{3\Delta P+4}-1)Q^2}{(1-\zeta^2)^2(\Delta P+1)}
\right)^{3/2}} .
\end{equation}
Expanding around the critical point using the critical quantities from Eq.~\eqref{P_cv_CT_c}, we obtain
\begin{eqnarray}
T_W = T_c + \frac{3}{8}\Delta P\, T_c
- \frac{3}{64}(\Delta P)^2 T_c + \mathcal{O}((\Delta P)^3) .
\end{eqnarray}
Defining $\Delta T = (T_W - T_c)/T_c$, we get
\begin{eqnarray}
\Delta T = \frac{3}{8}\Delta P
- \frac{3}{64} (\Delta P)^2
+ \frac{17}{1024} (\Delta P)^3
+ \mathcal{O}((\Delta P)^4) .
\end{eqnarray}
Inverting the series yields
\begin{equation}\label{dP_dT_Widom}
\Delta P = \frac{8}{3}\Delta T
+ \frac{8}{9}(\Delta T)^2
- \frac{20}{81}(\Delta T)^3 + \cdots .
\end{equation}
Thus, near criticality, the Widom line exhibits the universal mean-field behavior,
\[
\Delta P \propto \Delta T \ ,
\]
while higher-order terms encode curvature away from the linear regime. The monopole parameter $(1-\zeta^2)$ absorbed in the critical quantities $(v_c, T_c, P_c)$ and preserves the mean-field universality.


\subsection{Widom line in the canonical ensemble}

For the canonical ensemble, it is useful to express the variance $\Omega$ in terms of the entropy $S$. Using Eq.~\eqref{Temp_relation} and the relation between $r_+$ and $S$, Eq.~\eqref{eq:Variance} becomes
\begin{equation}
\Omega(S,Q)= \frac{8\pi^{3/2} L^2\sqrt{(1-\zeta^2) S^3}}{\pi L^2 (3\pi Q^2 - (1-\zeta^2)S) + 3S^2} .
\end{equation}
The Widom line at fixed charge is obtained from $\partial_S\Omega = 0$, giving
\begin{equation}\label{S_Widom}
S_W =
\frac{\pi}{6}
\left(
\sqrt{(1-\zeta^2)^2 L^4 + 108 L^2 Q^2}
- (1-\zeta^2)L^2
\right) .
\end{equation}
Substituting this into the temperature expression~\eqref{Temp_in_S_Q} yields
\begin{eqnarray}
T_W(S,Q)
= \frac{
12\sqrt{6}Q^2
}{
\pi \sqrt{1-\zeta^2}
\left(
\sqrt{(1-\zeta^2)^2 L^4 + 108 L^2 Q^2}
- (1-\zeta^2)L^2
\right)^{3/2}} .
\end{eqnarray}

With $\Delta Q = Q/Q_c - 1$, the expansion becomes
\begin{eqnarray}
T_W(S,Q)
= \frac{\sqrt{2/3}}{\pi L}
\left[
\frac{\Delta Q}{2\sqrt{6}\pi L}
- \frac{7\Delta Q^2}{8\sqrt{6}\pi L}
\right]  \ .
\end{eqnarray}
Using the critical temperature in Eq.~\eqref{S_cQ_cT_c}, and defining $\Delta T = (T_W - T_c)/T_c$, we obtain the Widom line
\begin{equation}\label{Widom_Line_dQDT}
\Delta T = - \frac{1}{4}\Delta Q
- \frac{7}{16} \Delta Q^2 + \mathcal{O}(\Delta Q^3) \ .
\end{equation}
Again, the monopole parameters are absorbed, and the universality is respected.

The analytic Widom lines obtained in both ensembles showed the universality of the supercritical behavior. We have not yet considered the conjugate field, so the system exhibits a symmetric supercritical crossover. In the next section, we consider the conjugate field that breaks this symmetry and splits this line into two distinct branches, $L^{\pm}$, which provide a more refined description of the supercritical region on the near-critical dynamics. In the next section, we compute the supercritical crossover lines $L^{\pm}$ within the global monopole threaded black hole.


\section{Supercritical Crossover}\label{Sec:Supercritical Crossover}

The supercritical region of a black hole phase diagram exhibits a smooth crossover rather than a genuine thermodynamic phase transition. Although the first-order small/large black hole transition terminates at the critical point, its characteristic imprints persist beyond it in the form of rapid yet continuous variations of thermodynamic response functions. Near the vicinity of criticality, the discontinuity disappears, and the system undergoes a continuous supercritical crossover instead. These supercritical crossover lines serve as probes of how critical fluctuations and mean-field scaling behavior extend beyond the critical point, providing a clear method for characterizing fluid-like behavior in the supercritical regime of black hole thermodynamics.

To analyze this structure, we introduce an appropriate order parameter and expand the thermodynamic potentials near the critical point. In the extended phase space, the order parameter naturally arises from the density variable associated with the specific volume, allowing the equation of state to be recast into a Landau form. In the canonical ensemble, entropy and charge fluctuations play analogous roles to those of an order parameter and an external field. Despite the distinction between these ensembles, both descriptions lead to the same universal mean-field scaling, including the characteristic nonanalytic behavior of the crossover curves. In the next subsections, we perform for both ensembles. 

\subsection{Supercritical Crossover in Extended Ensemble}

In the extended ensemble, as we discussed above, the order parameter is a density variable. To achieve this, we begin by rewriting the black hole equation of state in terms of a density variable as the inverse of specific volume $v$, i.e., $\rho = v^{-1}$. Just for completeness, we rewrite the EOS~\eqref{Main_EOS} for global monopole black holes as
\begin{eqnarray}
    P(\rho,T) = \rho T - \frac{\rho^2}{2\pi} + \frac{2 e^{-\zeta} Q^2}{\pi} \rho^4 \ ,
\end{eqnarray}
with the last term reflecting the effects of charge and the monopole parameter. At the critical point, the familiar inflection conditions
\[
\left.\frac{\partial P}{\partial \rho}\right|_c = 0 \ ,
\qquad
\left.\frac{\partial^2 P}{\partial \rho^2}\right|_c = 0
\]
guarantee that the dominant nonlinearities of the equation of state are cubic in deviations of $\rho$ from its critical value. Working with $\rho$ has two main advantages: first, the expansion coefficients around the critical point follow directly from derivatives of the EOS, and second, the connection to mean-field universality is manifest, as the resulting series mirrors that of standard Van der Waals systems.

Near the critical point, the thermodynamic ehavior of the system can be effectively described within the framework of a Landau-type expansion. The relevant order parameter is identified as the density deviation 
\begin{eqnarray}
    m = \rho - \rho_c \ ,
\end{eqnarray}
which quantifies departures from the critical density and distinguishes between the two coexisting phases below the critical temperature, $T_c$. Its conjugate field is the pressure deviation $\delta P$, which plays a role analogous to the external magnetic field in magnetic systems. Expanding the EOS to leading order yields
\begin{eqnarray}
    \delta P \equiv P(\rho,T) - P(\rho_c,T) = A\, m \tau + B \, m^3 + \cdots \ ,
\end{eqnarray}
with 
\begin{eqnarray}\label{ABINExtended}
    \tau = T-T_c \qquad;\qquad A = \left.\frac{\partial^2 P}{\partial \rho \partial T}\right|_c \qquad;\qquad B = \frac{1}{6}\left.\frac{\partial^3 P}{\partial \rho^3}\right|_c \ .
\end{eqnarray}
This correspondence follows from the thermodynamic potential $\mathcal{F}(m, T)$, whose minimization yields the mean-field equation of state and the pressure deviation, thus acts as the driving field controlling the response of the order parameter $m$. The isothermal susceptibility is associated with density fluctuations. It quantifies how strongly the order parameter $m$ responds to changes in its conjugate thermodynamic field, the pressure deviation $\delta P$. 

The form of $\chi_T$ follows directly from the Landau expansion of the equation of state, analogous to the magnetic susceptibility, which is defined as 
\begin{equation*}
    \chi_T = \frac{\partial M}{\partial H}\Bigg|_T \ ,
\end{equation*} 
In the case of a chosen order parameter, the form of the isothermal susceptibility $\chi_T$ is
\begin{eqnarray}\label{Chi_T}
    \chi_T = \left(\frac{\partial m}{\partial \delta P}\right)_T = \frac{1}{A\tau + 3B m^2} \ .
\end{eqnarray}
This relation expresses how sensitively the order parameter responds to infinitesimal perturbations in its conjugate field at a fixed temperature. A divergence of $\chi_T$ at the critical point reflects the emergence of long-range correlations and critical fluctuations, whereas the finite maximum of $\chi_T$ above $T_c$ traces the Widom line, representing the locus of maximal response in the supercritical regime.

To compute the fluctuations, we aim to extremize Eq.~\eqref{Chi_T}. By extremizing, we have 
\begin{eqnarray}
    m_\pm = \pm \sqrt{\frac{A}{3B}\tau} \qquad ; \qquad \delta P_\pm = \pm \frac{4}{3\sqrt{3}} \frac{A^{3/2}}{B^{1/2}} \tau^{3/2} \ .
\end{eqnarray}
The branches $m_+$ and $m_-$ correspond to fluctuations toward higher or lower densities relative to $\rho_c$. For the monopole-charged black hole, using Eq.~\eqref{ABINExtended}, we can easily compute the expressions for $A$ and $B$ as
\[
A = 1 \qquad;\qquad B = \frac{2\sqrt{6}}{3\pi} Q e^{-\zeta/2} \ .
\]
Thus, the crossover lines take the explicit form
\begin{equation}
    m_\pm = \pm \sqrt{\frac{\pi \tau}{2\sqrt{6}\,Q e^{-\zeta/2}}} \qquad ; \qquad \delta P_\pm = \pm \frac{4}{3}\sqrt{\frac{\pi}{2\sqrt{6}Q e^{-\zeta/2}}}\tau^{3/2} \ . 
\end{equation}
Now, we rewrite this expression in the reduced variables, 
\begin{eqnarray}
   \Delta T = \frac{T}{T_c}-1 \qquad\text{and}\qquad \Delta P = \tfrac{P}{P_c}-1  \ .
\end{eqnarray}
The crossover line in terms of the reduced variable is 
\begin{eqnarray}
    m_\pm = \pm \sqrt{\frac{\pi T_c}{2\sqrt{6}\,Q\, e^{-\zeta/2}}}\;\Delta T^{1/2} \qquad;\qquad \delta P_\pm(\Delta T) = \pm \frac{4}{3}\sqrt{\frac{\pi}{2\sqrt{6}\,Q\, e^{-\zeta/2}}}\; T_c^{3/2}\;\Delta T^{3/2}\  .
\end{eqnarray}
Defining $\Delta P_\pm = \frac{\delta P_\pm}{P_c}$, the two crossover branches follow
\begin{eqnarray}\label{DeltaP_from_Meanfield}
    \Delta P_{\pm}(\Delta T) = \pm \frac{32\sqrt{6}}{27}\, \Delta T^{3/2} + \cdots \ .
\end{eqnarray}
where the nonanalytic term $\propto \Delta T^{3/2}$ is universal and does not depend on $Q$ or $\zeta$. This symmetry between the two branches is precisely the gravitational analog of mean-field scaling familiar from fluid systems.
\begin{figure}[h!]
\centering
\begin{tikzpicture}
  \begin{axis}[
    domain=0:0.82,
    samples=300,
    xlabel={$\Delta T$},
    ylabel={$\Delta P$},
    xmin=0, xmax=0.82,
    ymin=0, ymax=2.0,
    width=15cm, height=8.5cm,
    grid=both,
    grid style={gray!20, dashed},
    legend pos=north west,
    legend cell align=left,
    every axis legend/.append style={nodes={scale=0.9, transform shape}}
  ]

    \addplot[name path=base, blue, thick] {8/3*x};
    \addlegendentry{Widom Line}

    \addplot[name path=pp, red, very thick] {8/3*x + (32*sqrt(6))/27 * x^(3/2)};
    \addlegendentry{$L^+$}

    \addplot[name path=pm, green!60!black, very thick] {8/3*x - (32*sqrt(6))/27 * x^(3/2)};
    \addlegendentry{$L^-$}

    \addplot[orange, fill=orange, fill opacity=0.12] fill between[of=pp and pm];

    \addplot[mark=*, mark options={scale=0.6, fill=black}] coordinates {(0,0)};
  \end{axis}
\end{tikzpicture}
\caption{Supercritical crossover structure in the $P$-$T$ plane. The central blue line denotes the mean ehavior $P(T)=\tfrac{8}{3}T$, while the red and green curves, $P_\pm(T)$, represent the upper and lower bounds determined by the supercritical corrections. The shaded region indicates the crossover fan that emerges from the critical point, and its locus traces the analogue of the Widom line in the black hole phase diagram.}
\label{fig:widomlinePT}
\end{figure}
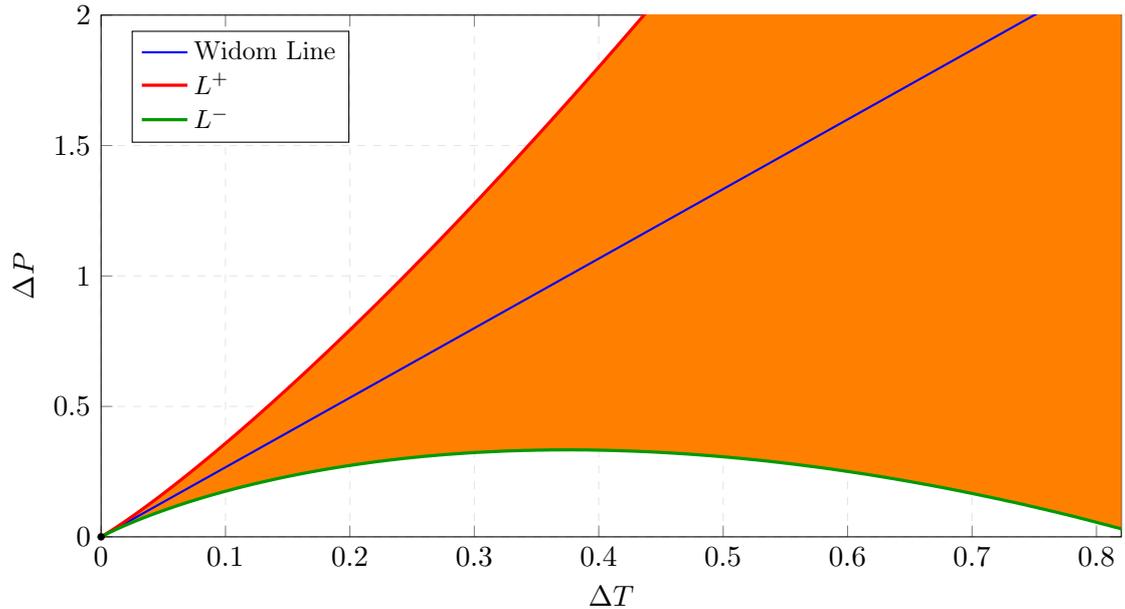

\begin{figure}[th]
    \centering
    \includegraphics[scale=0.49]{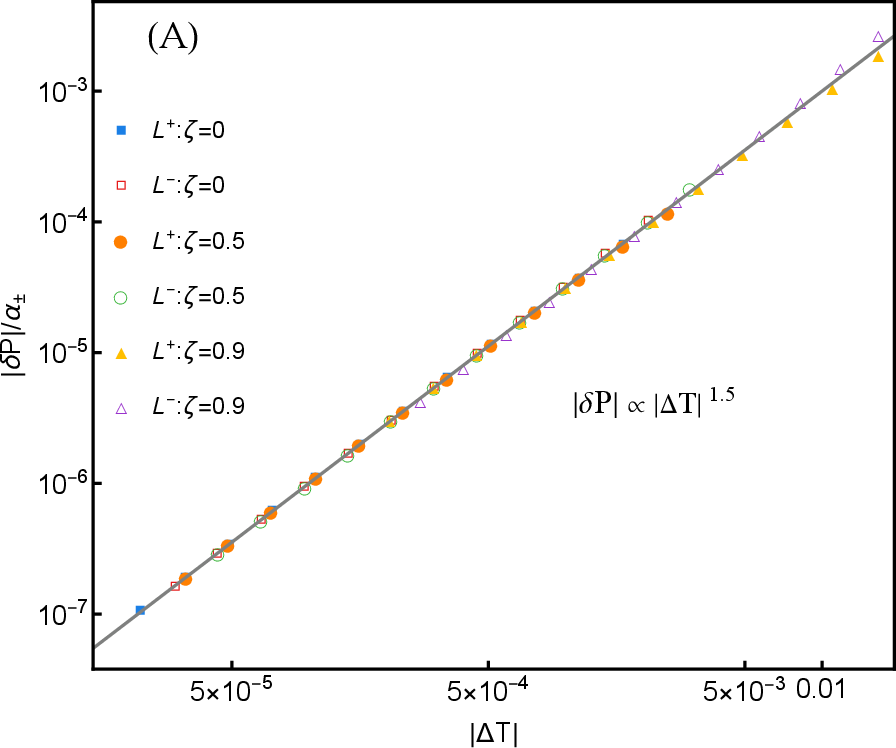}
    \includegraphics[scale=0.49]{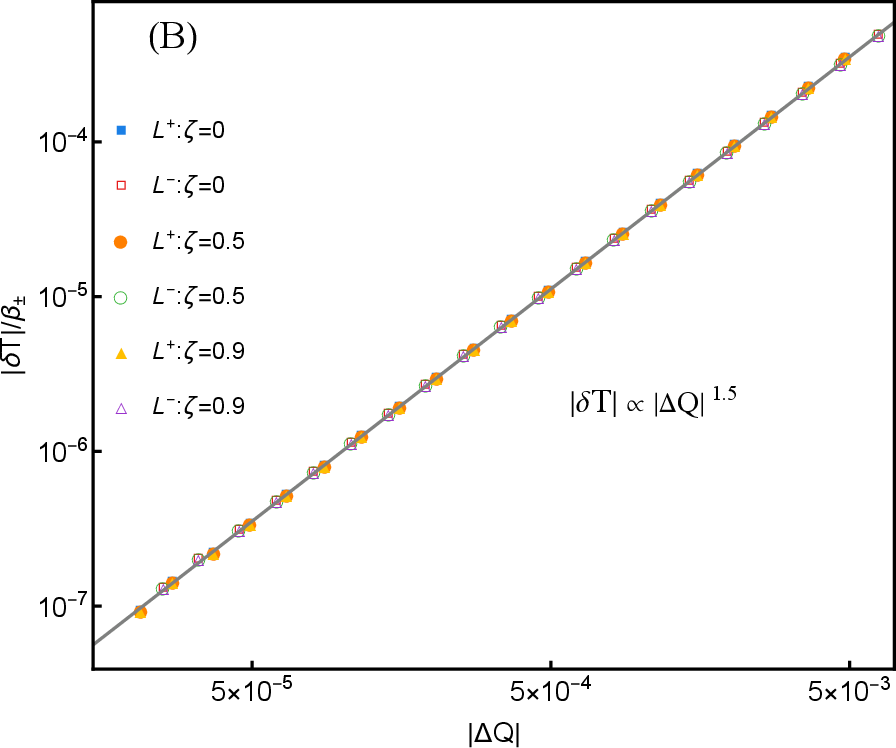}
    \caption{Scaling law of the ordering field $P$, along the supercritical crossover lines $L^{\pm}$ (filled markers and empty markers correspondingly) near the critical point, for the global monopole charged AdS black hole in the (A) extended phase space, (B) non-extended phase space (Canonical Ensemble). Here we include the data for three cases ($\zeta=0, 0.5, 0.9$). The solid lines represent power-law fits, with the exponents $\beta$ and $\gamma$ fixed by mean‑field universality (\(\beta = 1/2\), \(\gamma = 1\)), and the fitting parameters ($\alpha_\pm$ and $\beta_\pm$) are listed in Table~\ref{table:coefficients}.
 }
    \label{fig:scaling} 
\end{figure}

\begin{figure}[th]
    \centering
    \includegraphics[scale=0.42]{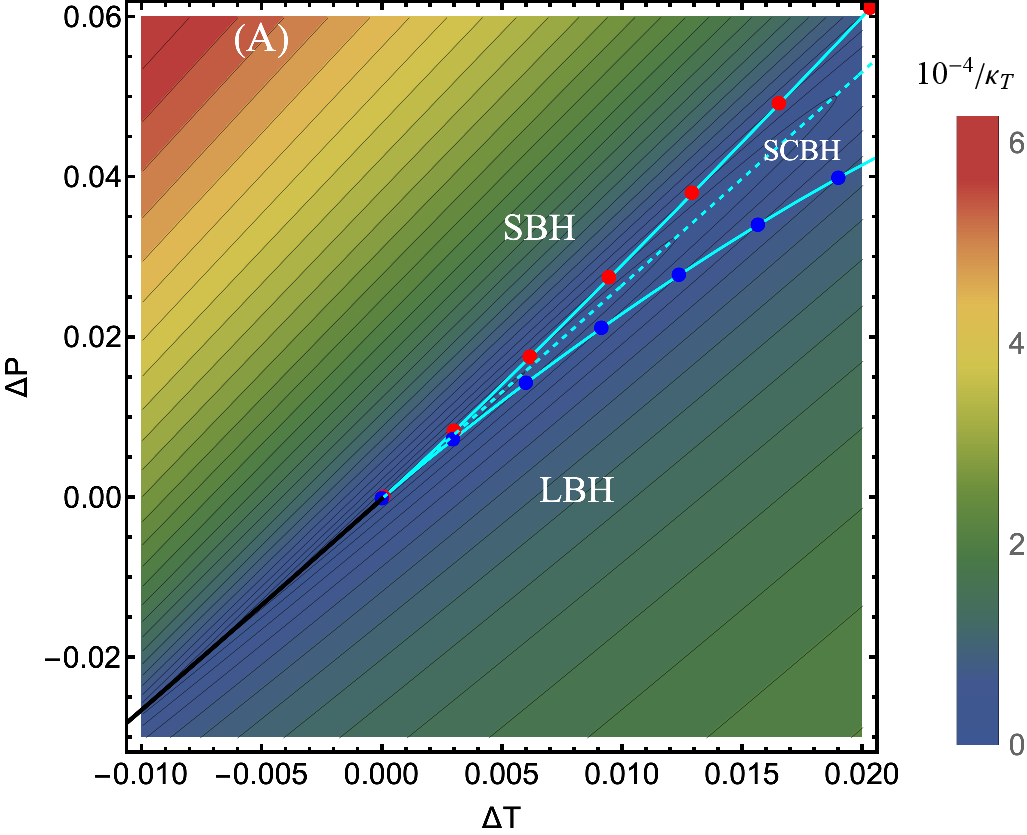}
    \includegraphics[scale=0.42]{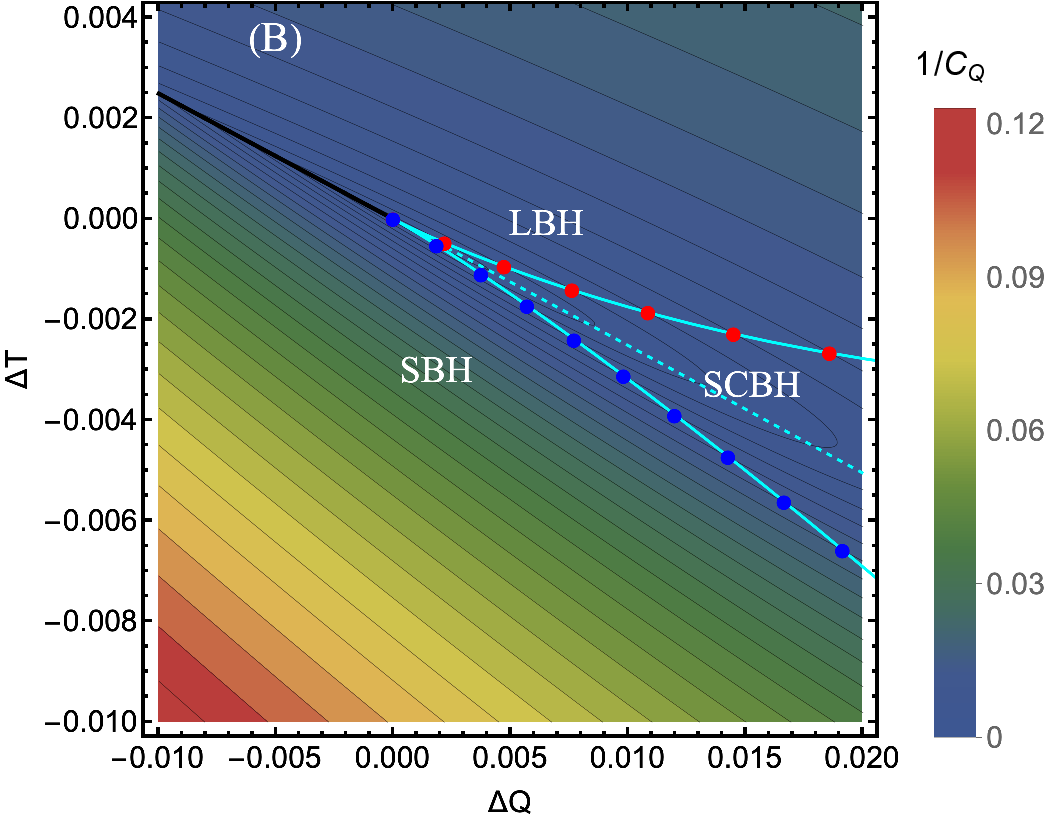}
    \caption{Phase diagrams corresponding to the global monopole charged AdS black hole in the (A) extended phase space, (B) non-extended phase space (Canonical Ensemble). The solid black and dashed cyan lines represent respectively the coexistence line and critical isochore. The solid black point marks the critical point. The solid cyan line with red (blue) points represents the $L^+$ ($L^-$) line. The color map and contour lines are obtained according to the corresponding thermodynamic response function indicated above the color bar. SBH: small black hole, LBH: large black hole, SCBH: SBH-LBH-indistinguishable (or supercritical BH). $\zeta=0.5$.
 }
    \label{fig:contour} 
\end{figure}

\begin{table}[!htbp]
\centering
\caption{Values of the coefficients $\alpha_\pm$ and $\beta_\pm$. }
\begin{tabular}{lrrrr}
\hline
\hline
\qquad \qquad & $ \alpha_+$ & $   \alpha_-  $ & $\beta_+$ & $\beta_- $\\
\hline
$\zeta=0$ & 0.0094 & 0.0098 & 0.1072 & 0.1019 \\
$\zeta=0.5$ & 0.0053 & 0.0055 & 0.1072 & 0.1019 \\
$\zeta=0.9$ & 0.00033 & 0.00037 & 0.1072 & 0.1019 \\
\hline
\hline
\end{tabular}
\label{table:coefficients}
\end{table}

Finally, by combining Eq.~\eqref{dP_dT_Widom} with the mean--field relation given in Eq.~\eqref{DeltaP_from_Meanfield}, we obtain the universal form of the crossover lines in the extended ensemble as  
\begin{eqnarray}
    \Delta P_{L^\pm}(\Delta T)  = \frac{8}{3}\,\Delta T  \;\pm\; \frac{32\sqrt{6}}{27}\,\Delta T^{3/2} + \cdots \ .
    \label{DeltaP_Lpm}
\end{eqnarray}
This expression shows the leading-order scaling behavior near the critical point. The linear term reflects the mean thermodynamic response of the system, while the nonanalytic $\Delta T^{3/2}$ contribution represents the dominant correction arising from critical fluctuations. The corresponding plot of $\Delta P_{L^\pm}(\Delta T)$ in Fig.~\ref{fig:widomlinePT} provides a clear visualization of the black hole equation of state in the supercritical regime. The central straight line, $P(T)$, traces the mean-field trajectory that characterizes the smooth evolution of the pressure with temperature. The nonlinear branches $P_\pm(T)$ form symmetric deviations about this mean, delineating the envelope of enhanced response. The region enclosed between these two branches defines the crossover fan, within which thermodynamic observables, such as specific heat, compressibility, and susceptibility, exhibit pronounced maxima. The central trajectory through this fan corresponds to the Widom line, marking the locus of maximal correlation length and providing a continuous extension of the first-order coexistence line into the supercritical domain.

To validate the theoretical scaling relations, we present numerical verification of the scaling law for the ordering field $P$ along the supercritical crossover lines $L^{\pm}$, which is based on the method introduced by \cite{Wang:2025ctk}, in Fig.~\ref{fig:scaling}(A) for the extended phase space. To define $L^{\pm}$ in the phase diagram, we first extend the coexistence line into the supercritical region by the critical isochore $v = v_c$. For a fixed pressure deviation $\delta P = P(v,T) - P(v_c,T)$, which corresponds to a path parallel to the critical isochore, we compute the isothermal compressibility $\kappa_T$ as a function of temperature. The temperature at which $\kappa_T$ attains its maximum along that path is denoted $T_{\text{max}}(\delta P)$. Collecting all such maxima for $\delta P > 0$ yields the $L^+$ branch, while those for $\delta P < 0$ give the $L^-$ branch. In the canonical (non‑extended) ensemble, the same construction is applied using the specific heat $C_Q$ as the response function and paths of fixed $\delta T = T(v,Q) - T(v_c,Q)$. It has been suggested that $L^{\pm}$ lines should follow the universal scaling law~\cite{li2024thermodynamic}:
 \begin{equation}
\delta P^{\pm} \propto (T-T_{\rm c})^{\beta + \gamma}\,,
\end{equation}
or 
\begin{equation}
\delta T^{\pm} \propto (Q-Q_{\rm c})^{\beta + \gamma}\,,
\end{equation}
near the critical control field $T_c$ or $Q_c$.

In Fig.~\ref{fig:scaling}(A), we consider three cases ($\zeta=0$, 0.5 and 0.9). The numerical data (scatters denoted by filled markers and empty markers for $L^+$ and $L^-$ respectively) are in excellent agreement with the power-law fits (solid lines), where the exponents are fixed by the mean-field universality class (\(\beta = 1/2\), \(\gamma = 1\)), and the coefficients $\alpha_{\pm}$ (listed in Table~\ref{table:coefficients}) are obtained from fitting. The results clearly show that the critical exponents (e.g., the $\Delta T^\frac{3}{2}$ scaling of the crossover branches) remain unchanged, while only the prefactors ($\alpha_\pm$), which encode the nonuniversal critical parameters, vary with $\zeta$. This confirms that the monopole parameter does not alter the universality class of the phase transition, in full agreement with our analytical mean‑field analysis. Furthermore, the phase diagram in terms of reduced variables ($\Delta T=T/T_c-1$ and $\Delta P=P/P_c-1$) is shown in Fig.~\ref{fig:contour}(A), which displays the coexistence line, critical isochore, and the $L^{\pm}$ lines, along with a color map of the corresponding thermodynamic response function. These lines originate at the critical point and smoothly divide the supercritical region into distinct thermodynamic zones. The reduced phase diagram in Fig.~\ref{fig:contour} appears unchanged under variation of $\zeta$, this is a consequence of our normalization by the critical parameters. In physical units, increasing $\zeta$ from 0 to 0.9 (corresponding to Fig.~\ref{fig:scaling}) dramatically shrinks the phase diagram: the critical point $(T_c, P_c)$ moves toward the origin, and the entire coexistence curve, Widom line, and $L^\pm$ branches are compressed into a smaller region of the $P-T$plane. Correspondingly, the isothermal compressibility $\kappa_T$ exhibits sharper peaks concentrated near the $L^\pm$ lines, reflecting the enhanced critical fluctuations as $\zeta$ approaches unity. In the limit $\zeta \rightarrow 1$, the critical temperature and pressure vanish, and the phase diagram collapses to a point, indicating that the small/large black hole transition is completely suppressed by the extreme solid angle deficit. Since the reduced phase diagram itself shows no discernible $\zeta-$dependence, we have chosen to present only the representative case $\zeta=0.5$ in Fig.~\ref{fig:contour} for clarity. These figures confirm the universal crossover structure derived analytically. This construction captures the universal features of black hole criticality in direct analogy with ordinary fluid systems near the liquid-gas critical point (see \cite{Wang:2025ctk} for details).

\subsection{Supercritical Crossover in Canonical Ensemble}\label{Sec:Canonical}

In this subsection, we focus on computing the supercritical crossover lines in the canonical ensemble, where the fluctuating thermodynamic variables are $S$ and $Q$. In this ensemble, the natural conjugate pair is the entropy $S$ and the charge $Q$, directly analogous to an order parameter responding to an external field in conventional mean-field systems. Near the critical point, we introduce small deviations
\begin{equation}
n = S - S_c \qquad;\qquad  \varepsilon = Q - Q_c \ ,
\end{equation}
representing fluctuations in entropy and charge. The temperature can be expanded as a Landau-type series in powers of $n$ near the criticality as
\begin{equation}
\delta T = J\,\varepsilon\,n + K\,n^3 + \cdots \ ,
\end{equation}
where
\begin{equation}
J = \left.\frac{\partial^2 T}{\partial S \partial Q}\right|_c = \frac{9\sqrt{6}}{2\pi^2 L^4} e^{-\zeta/2}  \qquad;\qquad K = \frac{1}{6}\left.\frac{\partial^3 T}{\partial S^3}\right|_c = \frac{9\sqrt{6}}{2\pi^4 L^7} \ .
\end{equation}
The bilinear term represents the linear coupling between charge and entropy fluctuations, while the cubic term captures nonlinear effects responsible for the bifurcation of phases near criticality, analogous to the double-well structure of Landau theory in condensed matter.

The susceptibility conjugate to charge fluctuations is
\begin{equation}
\chi_Q = \left(\frac{\partial n}{\partial \delta T}\right)_Q
= \frac{1}{J\varepsilon + 3K n^2} \ .
\end{equation}
The extrema of $\chi_Q$ indicate regions of maximal response, analogous to peaks in specific heat or compressibility in ordinary fluids. By solving for the extrema as we have done in the extended ensemble, we have
\begin{equation}
n_\pm^2 = \frac{J}{3K}\varepsilon  \qquad;\qquad \delta T_\pm = \pm \frac{4}{3\sqrt{3}}\,J^{3/2}K^{-1/2}\,\varepsilon^{3/2} \ .
\end{equation}
These define two distinct crossover branches, $L^+$ and $L^-$, which merge at the critical point, forming the Widom line structure that separates gas-like and liquid-like black hole phases. The scaling $\delta T_\pm \sim \varepsilon^{3/2}$ reflects mean-field universality.

\begin{figure}[h!]
\centering
\begin{tikzpicture}
  \begin{axis}[
    domain=0:0.71,
    samples=300,
    xlabel={$\Delta Q$},
    ylabel={$\Delta T$},
    xmin=-0.5, xmax=0.61,
    ymin=-0.5, ymax=0.2,
    width=15cm, height=8.5cm,
    grid=both,
    grid style={gray!20, dashed},
    legend style={at={(0.02,0.02)}, anchor=south west, font=\small}, 
    legend cell align=left
  ]

    \addplot[blue, thick] {-0.25*x};
    \addlegendentry{Widom Line}

    \addplot[name path=tp, green!70!black, very thick, domain=0:0.61] {-0.25*x + (1/sqrt(2))*x^(3/2)};
    \addlegendentry{$L^+$}
    \addplot[name path=tm, red, very thick, domain=0:0.61] {-0.25*x - (1/sqrt(2))*x^(3/2)};    
    \addlegendentry{$L^-$}

    \addplot[name path=tpL, green!70!black, very thick, domain=-0.5:0] {-0.25*x + (1/sqrt(2))*(-x)^(3/2)};
    \addplot[name path=tmL, red, very thick, domain=-0.5:0] {-0.25*x - (1/sqrt(2))*(-x)^(3/2)};

    \addplot[orange, fill=orange, fill opacity=0.12] fill between[of=tp and tm];
    \addplot[orange, fill=orange, fill opacity=0.12] fill between[of=tpL and tmL];

    \addplot[mark=*, mark options={scale=0.6, fill=black}] coordinates {(0,0)};
    
  \end{axis}
\end{tikzpicture}
\caption{Crossover behavior in the canonical ensemble. The central blue line represents the mean linear response, $\Delta T = -\tfrac{1}{4}\Delta Q$, while the upper and lower branches, $L^\pm$, encode nonlinear fluctuations scaling as $|\Delta Q|^{3/2}$. The shaded region forms the crossover fan, indicating the locus of maximal response in thermodynamic observables. This structure is the canonical analogue of the Widom line, smoothly connecting gas-like and liquid-like black hole phases in the supercritical regime.}
\label{fig:canonical-widom}
\end{figure}
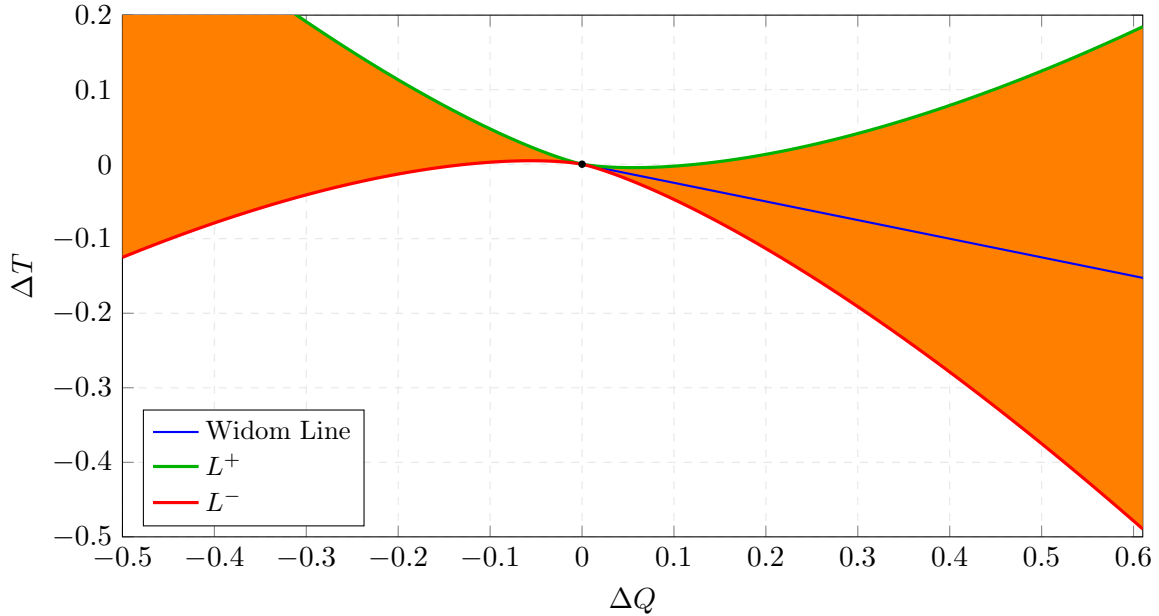

Again we introduce the reduced variables $\Delta T = \tfrac{T}{T_c}-1$ and $\Delta Q = \frac{Q}{Q_c}-1$ the universal form of the crossover lines is
\begin{equation}
\Delta T_\pm = \frac{\delta T_\pm}{T_c} = \pm \frac{1}{\sqrt{2}}\,|\Delta Q|^{3/2} \ .
\end{equation}
All explicit dependence on model parameters cancels, confirming the universality of the critical behavior. Finally, the complete expression for the canonical Widom lines using Eq.~\eqref{Widom_Line_dQDT} reads
\begin{equation}
\Delta T_{L^\pm}(\Delta Q) = -\frac{1}{4}\,\Delta Q 
\;\pm\; \frac{1}{\sqrt{2}}\,|\Delta Q|^{3/2} + \cdots  \ .
\end{equation}
The linear term represents the average thermodynamic trajectory near criticality, while the nonlinear correction encodes symmetric fluctuations that split the response into two crossover branches. The canonical ensemble exhibits the same universal crossover behavior as the extended phase space. The linear drift of the coexistence curve and the nonlinear splitting are consistent with standard mean-field critical phenomena, highlighting the deep connection between black hole thermodynamics and classical liquid-gas transitions or magnetic systems. This can also be seen pictorially in Fig.~\ref{fig:canonical-widom}. Fig.~\ref{fig:canonical-widom} analytically illustrates the canonical ensemble crossover near the critical point. The central blue line represents the linear mean-field response, \(\Delta T = -\tfrac{1}{4}\Delta Q\), while the branches \(L_+\) and \(L_-\) depict nonlinear fluctuations scaling as \(|\Delta Q|^{3/2}\). The shaded region between them forms the crossover fan, indicating the locus of maximal response in thermodynamic observables. This fan represents the canonical analogue of the Widom line, smoothly connecting gas-like and liquid-like black hole phases in the supercritical regime and highlighting the universal mean-field critical behavior.

For the canonical ensemble, the scaling law of $\delta T$ versus $\Delta Q$ along $L^{\pm}$ is numerically verified in Fig. \ref{fig:scaling}(B), where the data points align with the theoretical power-law fits, affirming the universal exponents. It is noteworthy that the fitting parameters $\beta_\pm$ are independent of the value of $\zeta$ in the non-extended phase space. The corresponding phase diagram in reduced variables is presented in Fig. \ref{fig:contour}(B), illustrating the coexistence line, critical isochore, and the $L^{\pm}$ lines, complemented by a color map of the thermodynamic response. These figures provide a comprehensive view of the crossover structure in the canonical setting.

Importantly, the phase structure revealed in Fig. \ref{fig:contour} indicates that in the supercritical region, the global monopole charged AdS black hole in the extended phase space (panel (A)) behaves analogously to a liquid-gas system, while in the canonical ensemble (panel (B)) it resembles a liquid-liquid phase transition system, consistent with the water phase diagrams as shown in~\cite{Wang:2025ctk}.

\section{Summary and Outlook}\label{Sec:Conclusion}

In this work, we presented a fully analytical investigation of the Widom line and the supercritical crossover structure of charged AdS black holes with a global monopole. Starting from the monopole-modified geometry, we studied the thermodynamics in both the extended and canonical ensembles, showing explicitly how the solid-angle deficit rescales the horizon geometry and shifts all critical thermodynamic variables. By computing the normalized variance $\Omega$ from the Gibbs free energy and performing a Landau-type expansion near criticality, we derived closed-form expressions for the Widom line and the bifurcating crossover branches $L^{\pm}$ in both ensembles.

Our results demonstrate that while the monopole parameter quantitatively shifts the critical point and modifies the slopes and curvature of the supercritical trajectories, but it does not alter the underlying universality class. The leading linear relation $\Delta P \propto \Delta T$ and the characteristic nonanalytic mean-field correction $\Delta P \propto \Delta T^{3/2}$ in the extended ensemble and $\Delta T \propto \Delta Q$ and the nonanalytic mean-field correction $\Delta T \propto \Delta Q^{3/2}$ in the canonical ensemble remain universal. These analytic findings are supported by numerical verification: the scaling laws along the crossover branches, as well as the phase diagrams displaying coexistence lines, critical isochores, and the $L^{\pm}$ curves, all exhibit excellent agreement with the theoretical predictions. This reinforces the robustness of mean-field universal scaling across both ensembles.

Several promising avenues for further exploration remain open, and we will revisit these soon. By using Ruppeiner geometry or correlation-volume diagnostics to quantify how the monopole parameter modifies microscopic interactions and fluctuation patterns across the Widom line. Investigating quasi-normal modes, relaxation times, or dynamical susceptibilities to determine whether signatures of critical slowing down persist in the presence of topological defects. Interpreting monopole-induced thermodynamic deformations in the dual CFT, where the solid-angle deficit may correspond to defect backgrounds or modified vacuum structures. Extending the analysis to rotating black holes, Born-Infeld or higher-curvature theories, scalar-hairy solutions, or higher dimensions to examine the universality of the $\Delta T^{3/2}$ scaling under broader gravitational deformations. Studying Widom-line behavior in time-dependent or non-equilibrium settings, such as AdS black hole evaporation or holographic quenches, to explore dynamical crossover phenomena will yield promising results.


\section*{Acknowledgments}
 Ankit Anand is financially supported by the Institute's postdoctoral fellowship at the Indian Institute of Technology Kanpur.

\bibliographystyle{utphys.bst}
\bibliography{ref}
\end{document}